\documentclass[a4paper]{jpconf}
\usepackage{graphicx}

\usepackage{lineno}
\usepackage[figuresright]{rotating}

\usepackage[utf8]{inputenc}

\newcommand{\pp}{{\rm pp}}

\newcommand{\jpsi}{{\rm{J}/\psi}}
\newcommand{\psiprime}{{\psi{\rm(2S)}}}
\newcommand{\chic}{{\chi_{c}}}

\renewcommand{\ss}{{\sqrt{s}}}

\renewcommand{\pt}{{p_{\rm T}}}

\newcommand{\mc}{{m_c}}
\newcommand{\alphas}{{\alpha_{\rm s}}}

\begin{document}
\title{Charmonium production in $\pp$ collisions with ALICE at the LHC}

\author{Hugo Pereira Da Costa, for the ALICE Collaboration}

\address{Commissariat \`a l’Energie Atomique, IRFU, Saclay, France}

\ead{pereira@hep.saclay.cea.fr}

\begin{abstract}
We report on forward-rapidity charmonium production in pp collisions at a center of mass energy $\ss=13$ TeV, as measured by ALICE at the LHC. Differential cross sections for both $\jpsi$ and $\psiprime$ are presented as a function of the charmonium transverse momentum and rapidity. Results are compared to similar measurements performed by LHCb, to lower energy measurements and to state of the art model calculations. 
\end{abstract}

\vspace*{-3mm}
\section*{}
\label{}

Charmonia (e.g. $\jpsi$ and $\psiprime$) are mesons formed of a charm and anti-charm quark pair. In high-energy hadronic collisions, their production results mainly from the hard scattering of two gluons in a process characterized by a timescale $\sim 1/2 \mc = 0.08$ fm/$c$ (with $\mc$ the mass of the charm quark), followed by the hadronization of the charm quark pair in a charmonium state with a timescale $\sim 1/\alphas \mc = 0.6$ fm/$c$ (with $\alphas$ the strong coupling constant). While the production of the charm quark pair is relatively well described by perturbative QCD calculations~\cite{Cacciari:2012ny}, their binding into a charmonium state is inherently non-perturbative and poses a challenge to models~\cite{Brambilla:2010cs}. It is therefore important to measure the charmonium cross sections in proton-proton ($\pp$) collisions over large transverse momentum ($\pt$) and rapidity ($y$) ranges, as well as for as many collision energies as possible, in order to understand their production mechanism. Moreover these same measurements also provide a reference baseline for proton-nucleus and nucleus-nucleus measurements, which in turn are used to quantify the properties of the Quark-Gluon Plasma~\cite{Shuryak:1978ij}.

\begin{figure}[h]
\begin{center}
\begin{tabular}{cc}
\includegraphics[width=0.47\linewidth,keepaspectratio]{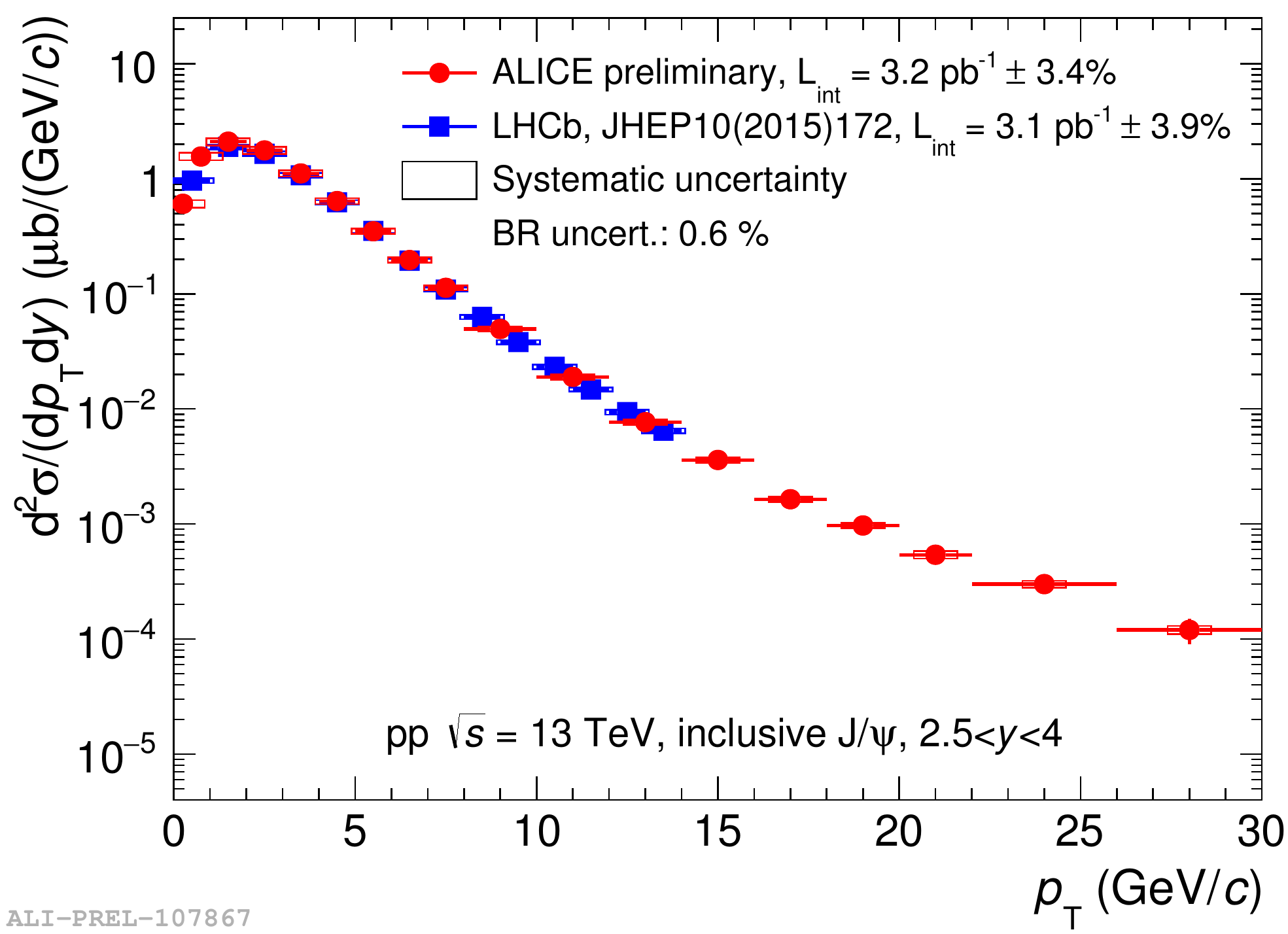}&
\includegraphics[width=0.47\linewidth,keepaspectratio]{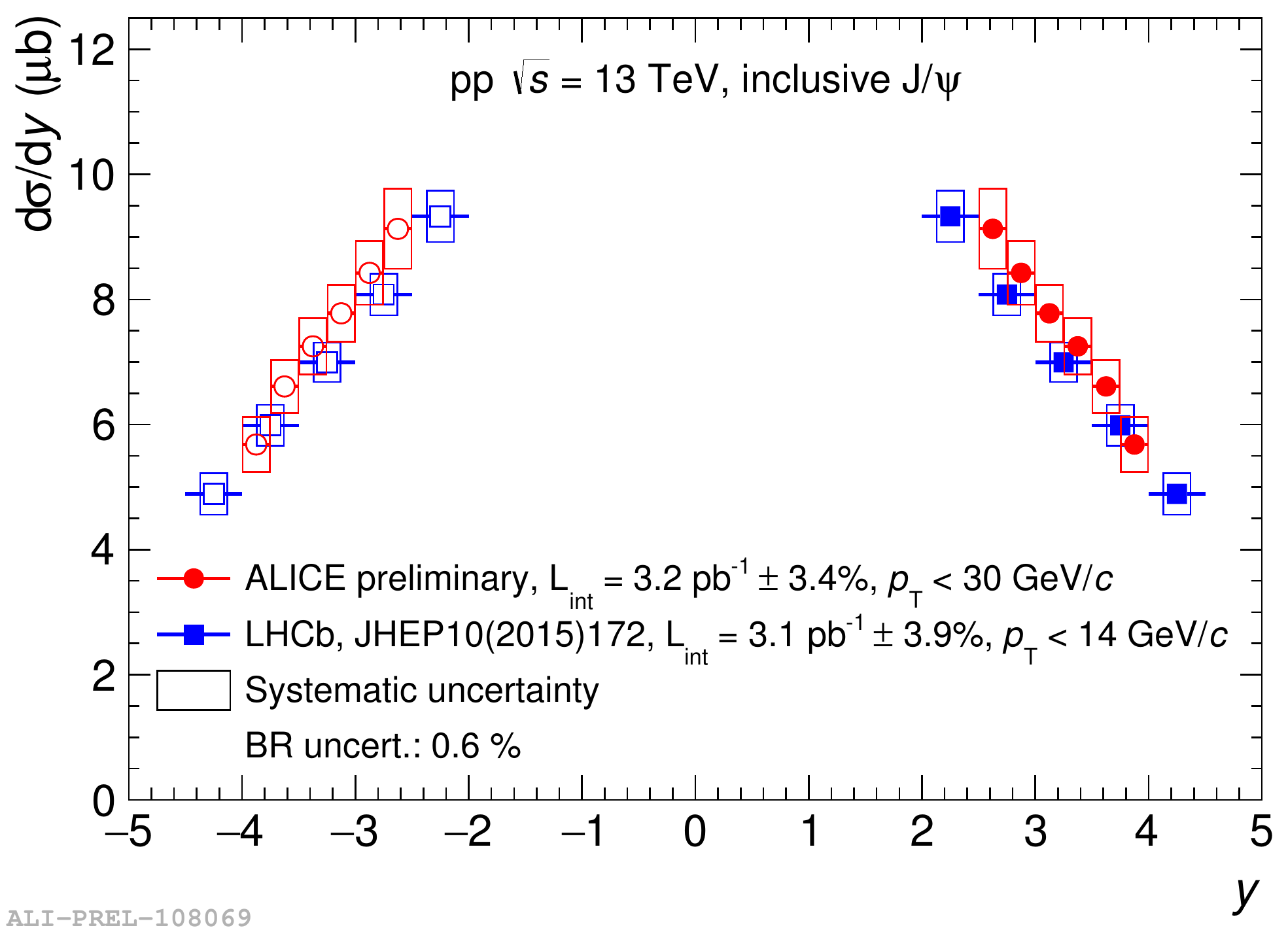}\\
\includegraphics[width=0.47\linewidth,keepaspectratio]{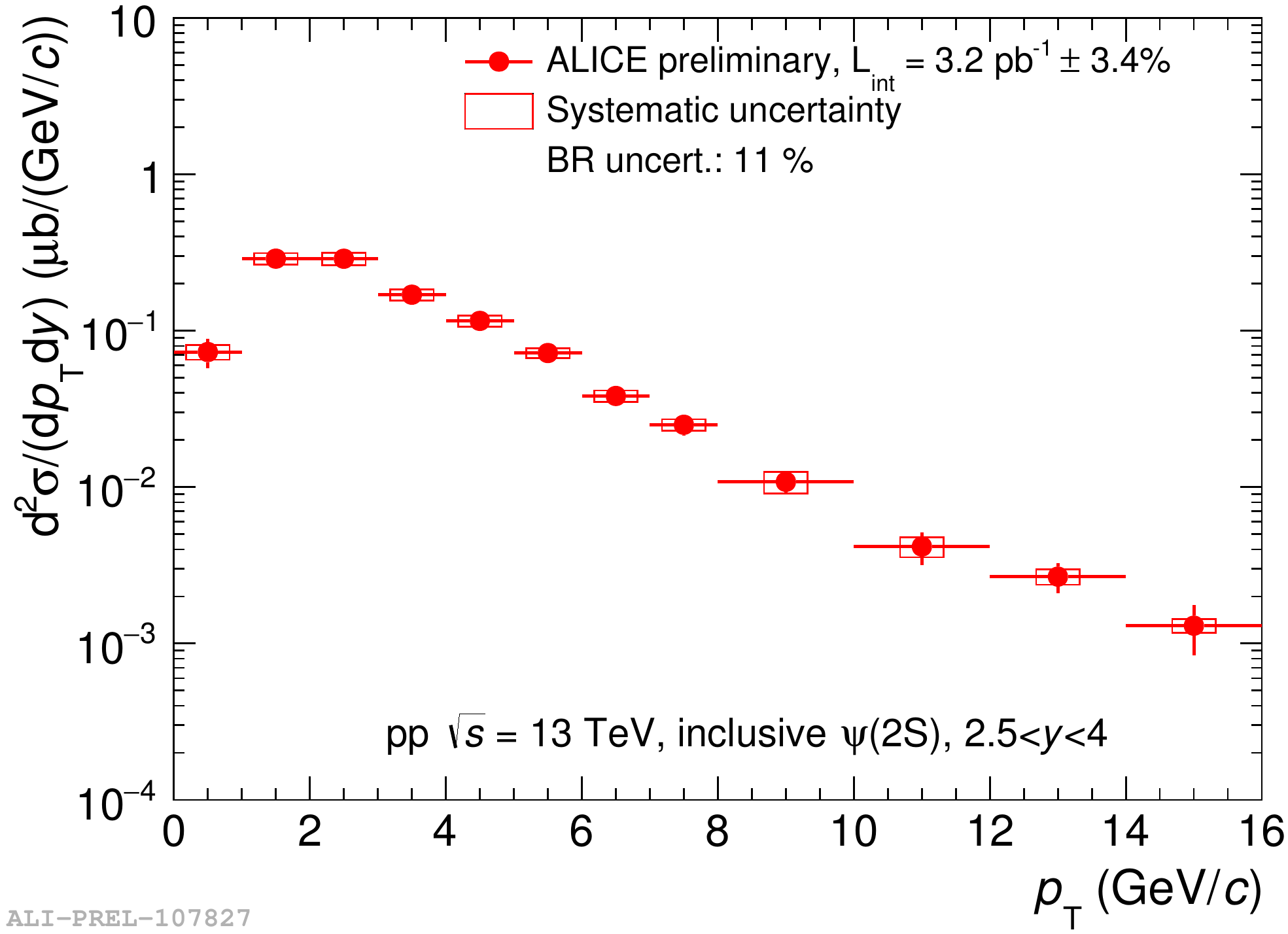}&
\includegraphics[width=0.47\linewidth,keepaspectratio]{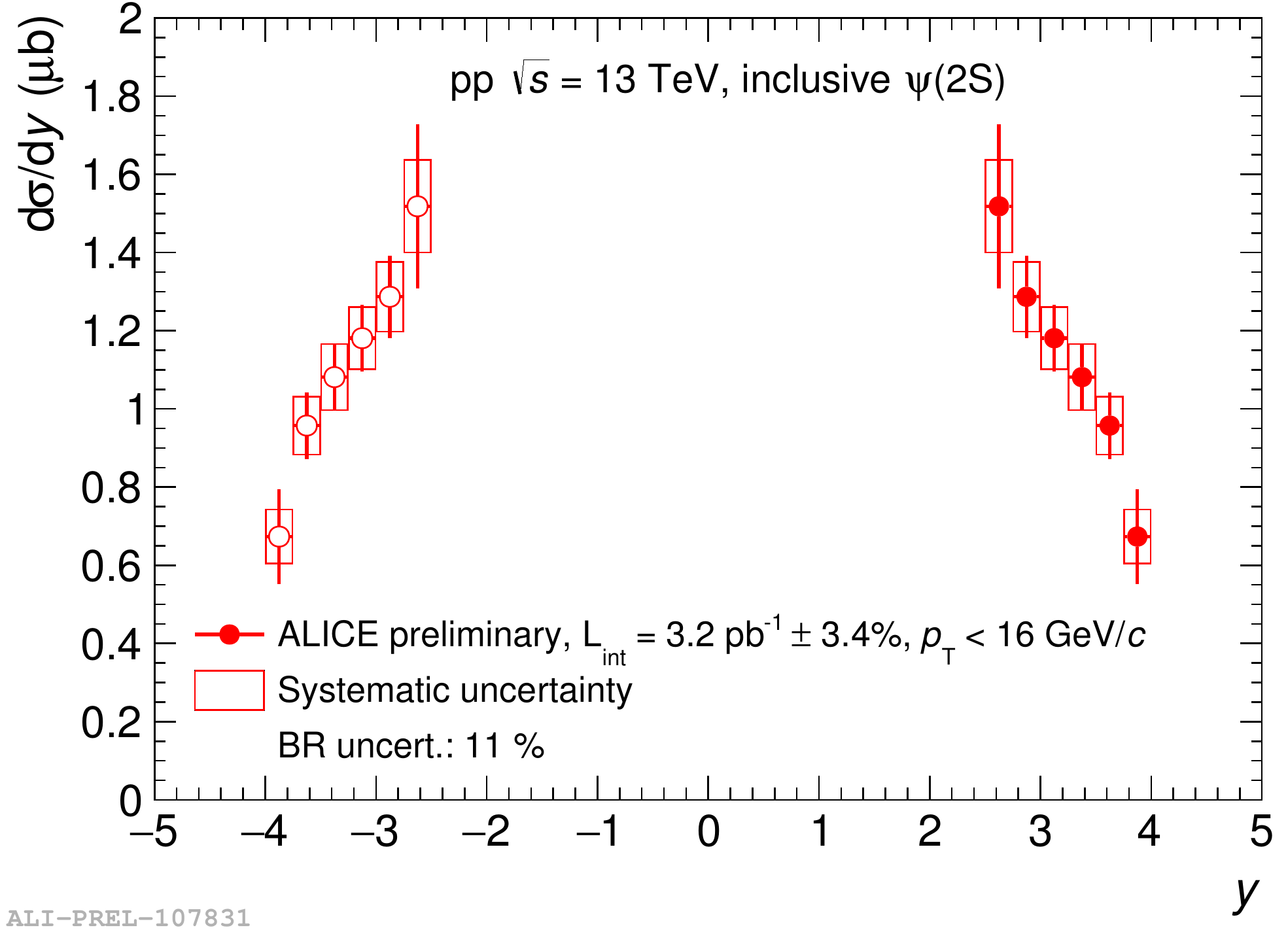}
\end{tabular}
\end{center}
\vspace*{-3mm}
\caption{\label{fig_sigma_jpsi} Inclusive forward-rapidity $\jpsi$ (top) and $\psiprime$ (bottom) cross sections as a function of $\pt$ (left) and $y$ (right) in $\pp$ collision at $\ss=13$~TeV.}
\end{figure}

Figure~\ref{fig_sigma_jpsi} shows the inclusive $\jpsi$ (top) and $\psiprime$ (bottom) production cross sections as a function of $\pt$ (left) and $y$ (right), measured by ALICE at forward rapidity ($2.5<y<4$) in $\pp$ collisions and at a center of mass energy $\ss=13$~TeV. The analysis technique used for these measurements is very similar to that of lower energy charmonium measurements performed by ALICE, as described for instance in~\cite{Adam:2015rta}. In particular, the number of charmonia is measured using fits to the invariant mass distribution of opposite sign muons detected in the ALICE Muon Spectrometer~\cite{ALICE,Aamodt:2011gj}. The data sample used for this analysis corresponds to an integrated luminosity $L=3.2$~pb$^{-1}$. Systematic uncertainties on the cross sections include contributions from signal extraction, acceptance and efficiency corrections as well as the integrated luminosity. They amount to about 7\% for $\jpsi$ and 10\% for $\psiprime$. For $\jpsi$, cross sections are compared to measurements done by LHCb in the same rapidity range~\cite{Aaij:2015rla}. An excellent agreement is observed between the two experiments. ALICE $\jpsi$ measurements extends the $\pt$ reach from $14$ to $30$~GeV/$c$. For $\psiprime$ this is the only measurement available at this energy and in this rapidity range.

All cross sections shown in figure~\ref{fig_sigma_jpsi} are inclusive. They consist of a prompt contribution that includes direct production and (for $\jpsi$) decay from $\psiprime$ and $\chic$, and a non-prompt contribution from $b$-hadron decays. Both the prompt and non-prompt contributions must be properly accounted for when comparing the data to calculations.

\begin{figure}[t]
\begin{center}
\begin{tabular}{cc}
\includegraphics[width=0.47\linewidth,keepaspectratio]{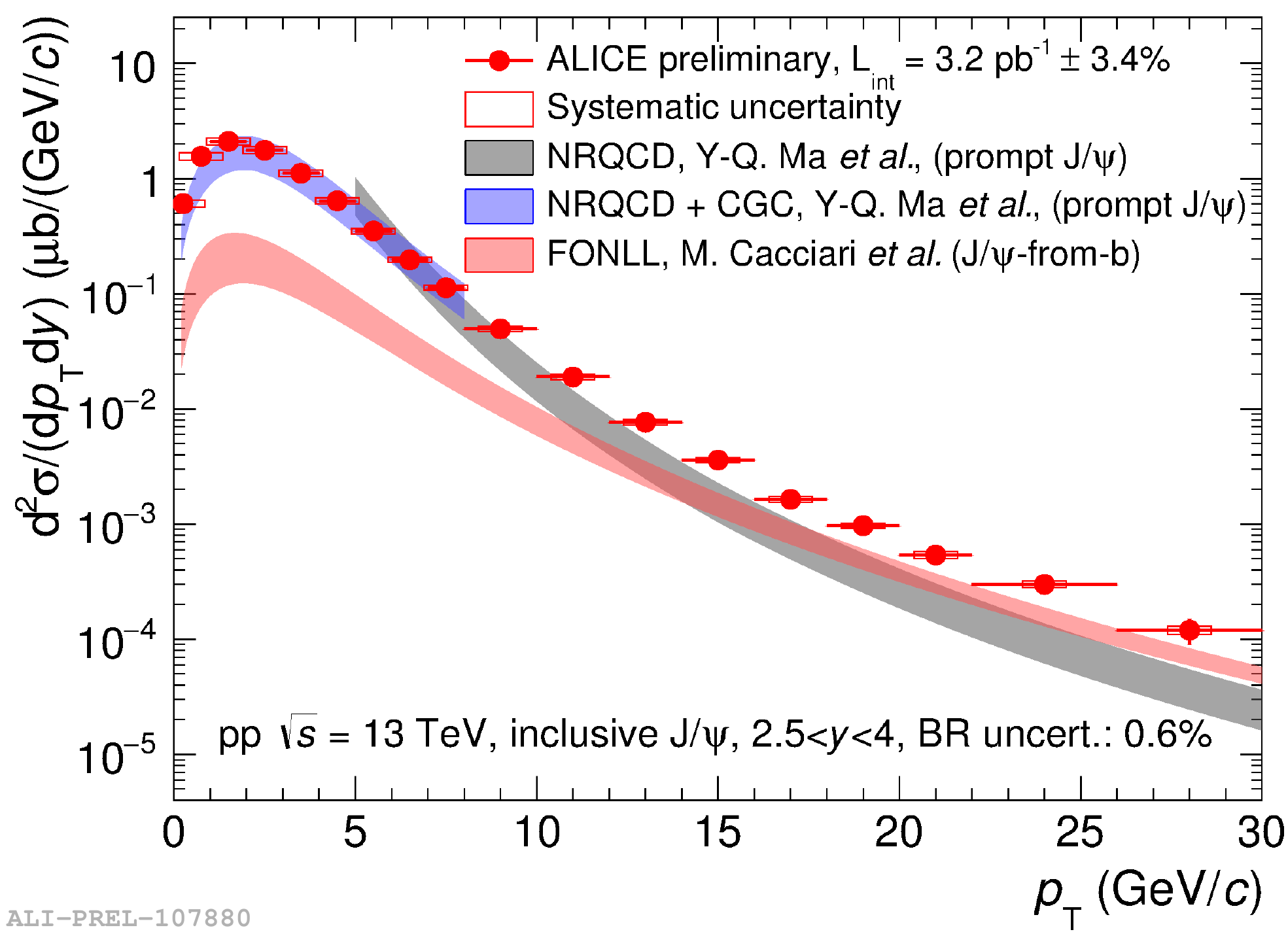}&
\includegraphics[width=0.47\linewidth,keepaspectratio]{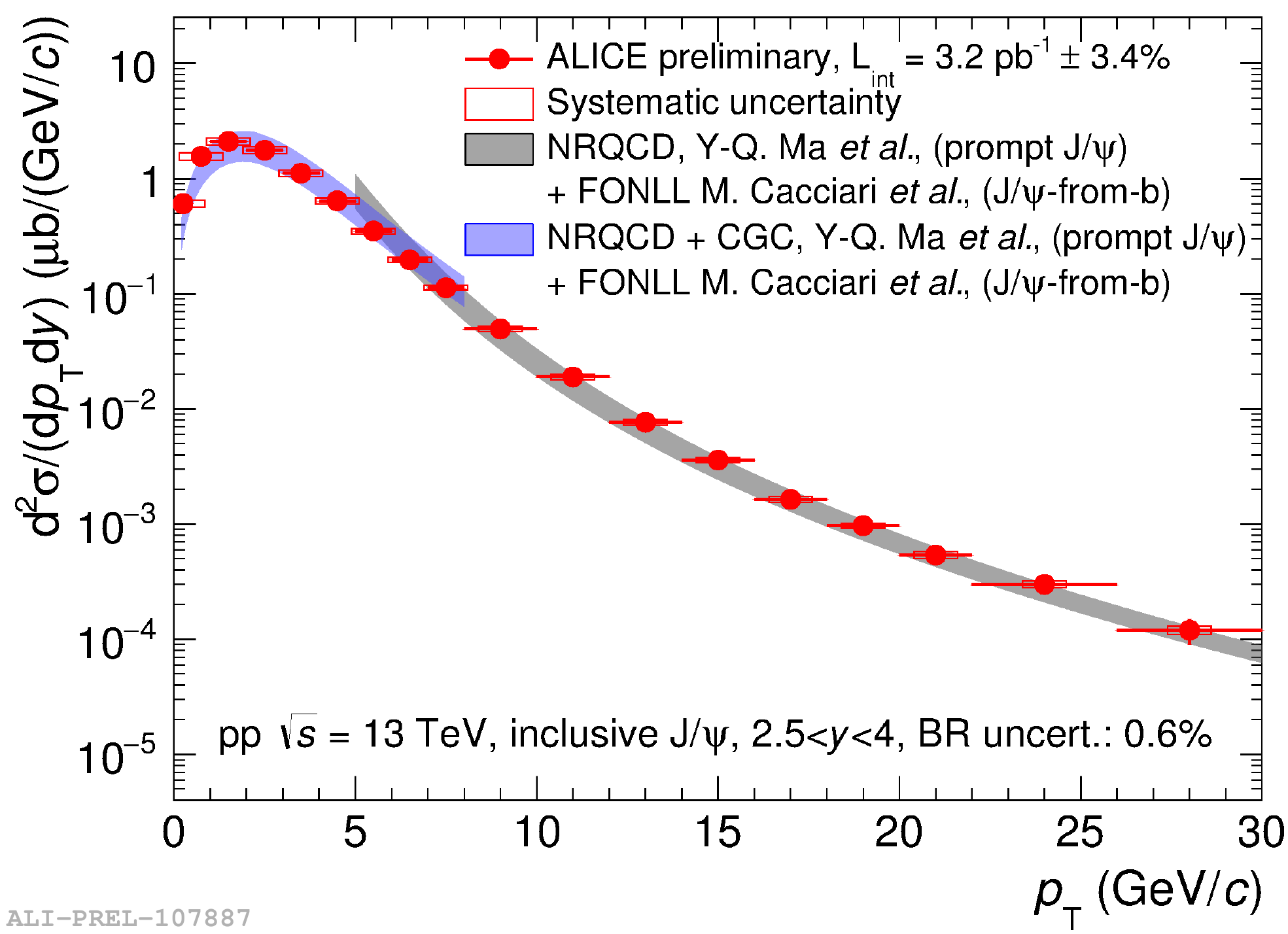}
\end{tabular}
\end{center}
\vspace*{-3mm}
\caption{\label{fig_jpsi_yqma_pt} Inclusive forward-rapidity $\jpsi$ cross section as a function of $\pt$ compared to model calculations. See text for details about the models.}
\end{figure}

One such comparison is shown in figure~\ref{fig_jpsi_yqma_pt}. The left panel shows the inclusive $\jpsi$ cross section as a function of $\pt$, together with three calculations: (i) at high $\pt$ in grey, a calculation of the prompt $\jpsi$ contribution, in the framework of Non-Relativistic QCD (NRQCD) at Next-to-Leading Order (NLO)~\cite{Ma:2010yw}, (ii) at low $\pt$ in blue, a Leading Order (LO) NRQCD calculation coupled to a Color Glass Condensate (CGC) description of the low-$x$ gluons in the proton~\cite{Ma:2014mri} and (iii) in red, a Fixed-Order Next-to-Leading Logarithm (FONLL) calculation of the non-prompt $\jpsi$ production~\cite{Cacciari:2012ny}. Based on these calculations, the non-prompt contribution to inclusive $\jpsi$ production is of order 10\% at low $\pt$, it increases steadily with increasing $\pt$ and dominates for $\pt$ larger than $\sim 20$~GeV/$c$.
At low $\pt$, where the non-prompt contribution is negligible, the NRQCD+CGC prompt $\jpsi$ calculation reproduces the data reasonably well. For $\pt>8$~GeV/c the NRQCD calculation is significantly below the data, which is expected due to the onset of the non-prompt contribution. 

In the right panel of figure~\ref{fig_jpsi_yqma_pt}, the calculations for prompt and non-prompt $\jpsi$ production are summed, assuming fully uncorrelated uncertainties. A good description of the data is obtained over the full $\pt$ range and spanning more than 4 orders of magnitude in cross sections. A similar level of agreement is also observed when using the NRQCD calculation from~\cite{Butenschoen:2010rq} instead of the ones shown here or when considering the inclusive production of the $\psiprime$ meson instead of $\jpsi$.

\begin{figure}[h]
\begin{center}
\begin{tabular}{cc}
\includegraphics[width=0.48\linewidth,keepaspectratio]{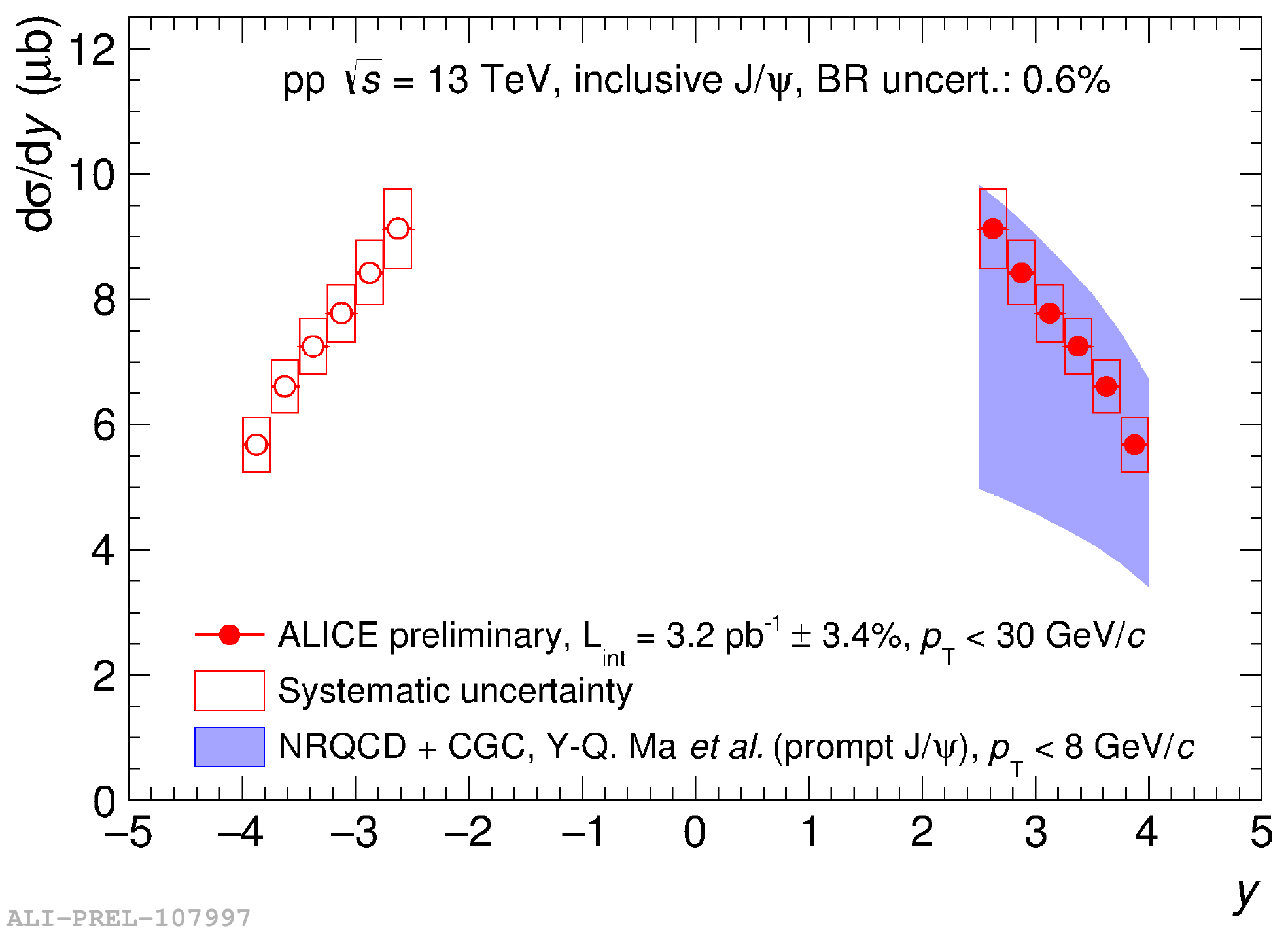}&
\includegraphics[width=0.48\linewidth,keepaspectratio]{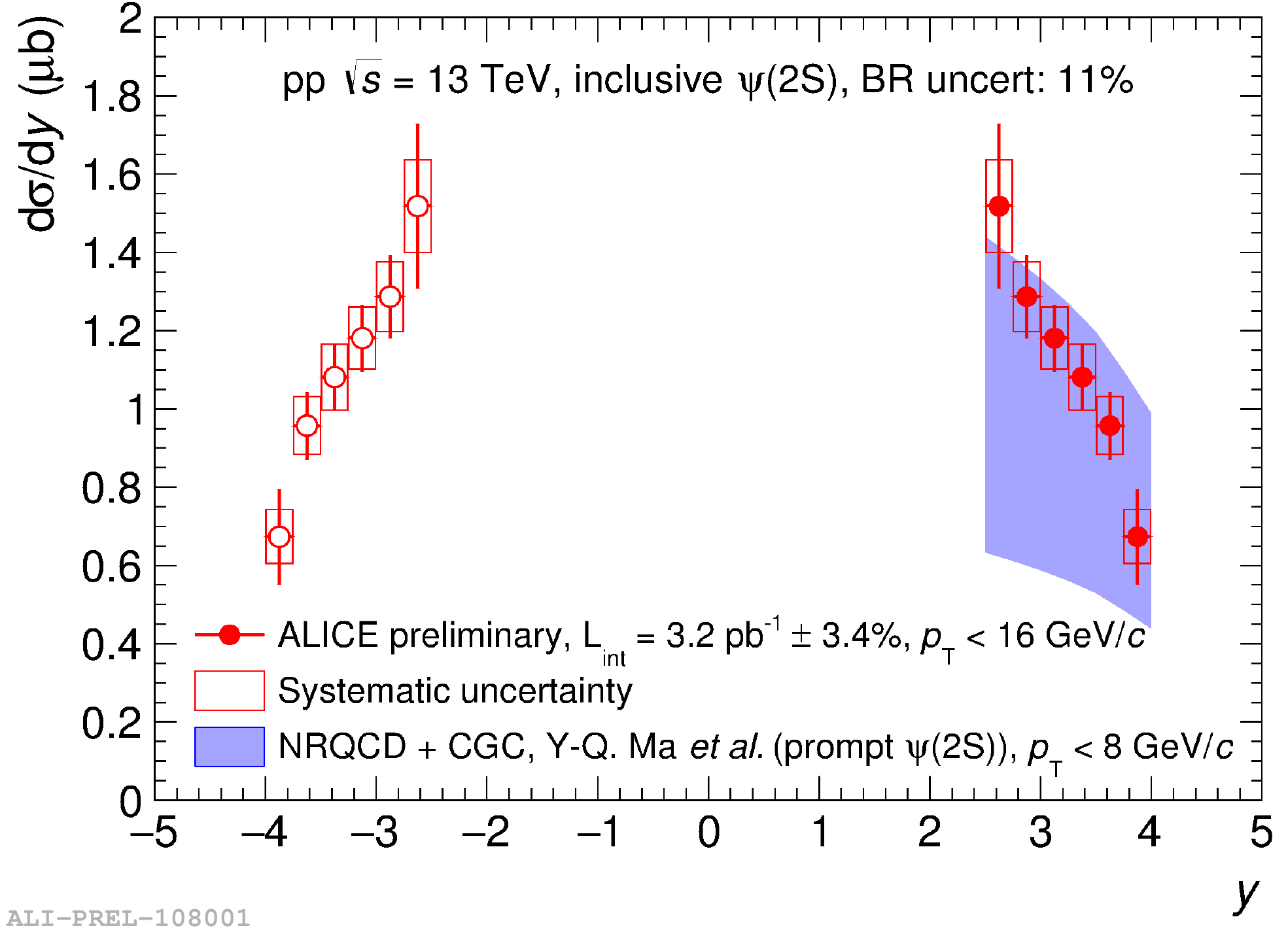}
\end{tabular}
\end{center}
\vspace*{-3mm}
\caption{\label{fig_jpsi_yqma_y} Inclusive forward-rapidity $\jpsi$ (left) and $\psiprime$ (right) cross section as a function of $y$ compared to a NRQCD + CGC model calculation.}
\end{figure}

Since the calculation from \cite{Ma:2014mri} extends down to $\pt=0$~GeV/$c$, it can be integrated over $\pt$ and compared to ALICE $y$-differential cross section measurements. A reasonable agreement is observed between model and data for both $\jpsi$ and $\psiprime$, as shown in figure~\ref{fig_jpsi_yqma_y}, although the theoretical uncertainties are rather large. This agreement would be further improved if one would add the non-prompt contribution to the model, which amounts to $10-15$\% according to FONLL. 

\begin{figure}[h]
\begin{center}
\begin{tabular}{cc}
\includegraphics[width=0.48\linewidth,keepaspectratio]{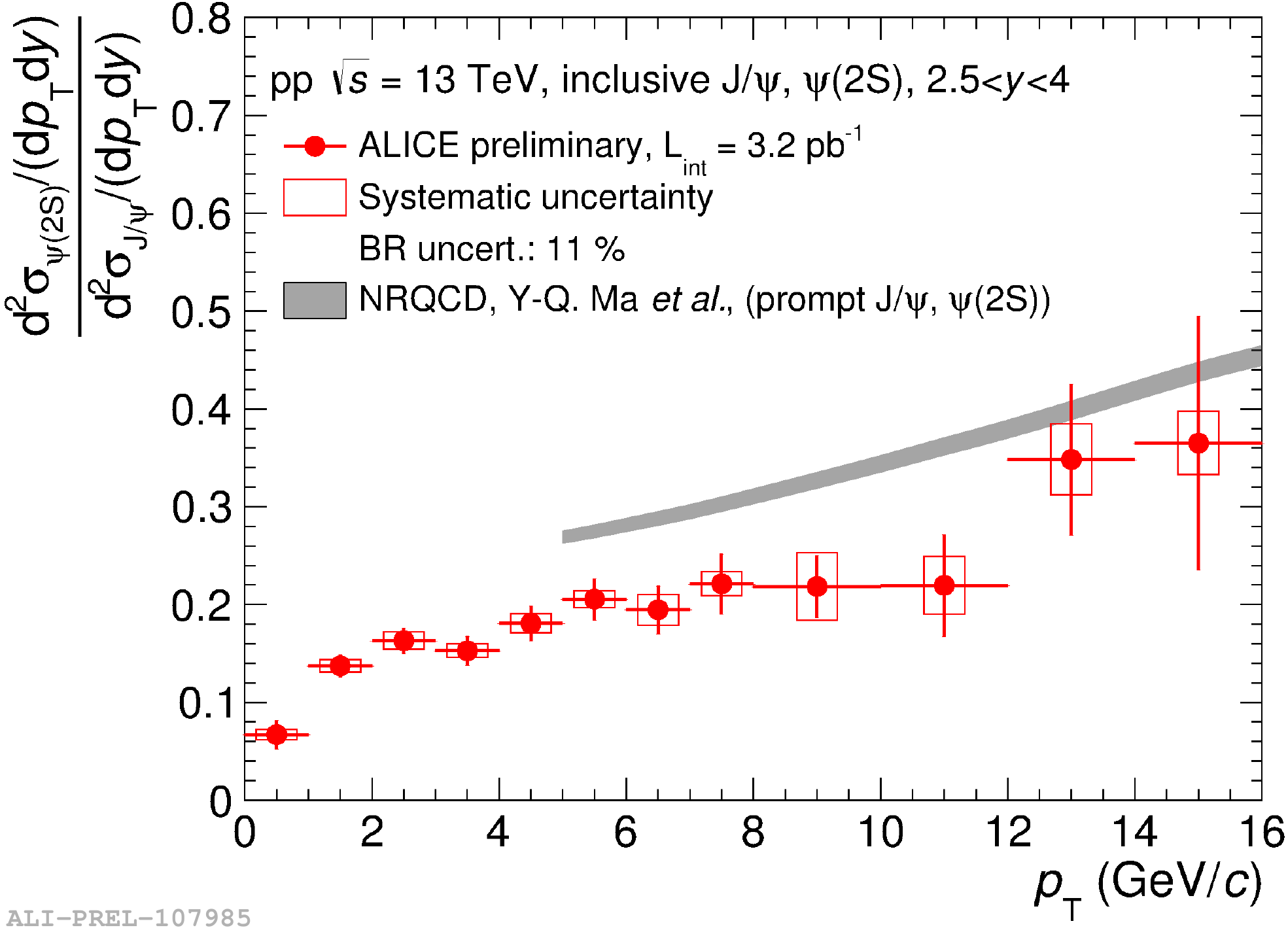}&
\includegraphics[width=0.48\linewidth,keepaspectratio]{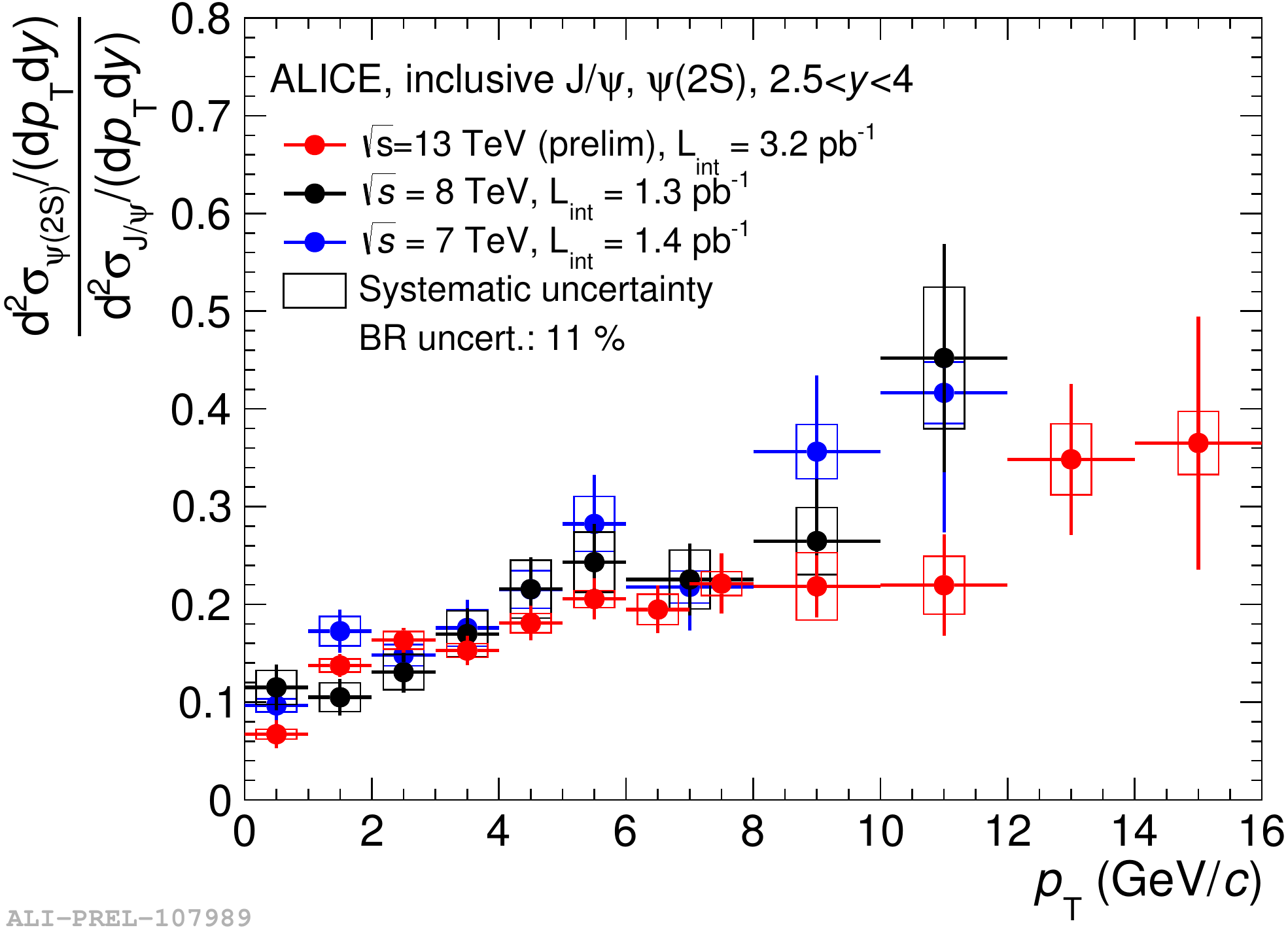}
\end{tabular}
\end{center}
\vspace*{-3mm}
\caption{\label{fig_ratio} $\psiprime$-to-$\jpsi$ inclusive cross section ratio as a function of $\pt$ in pp collisions at $\ss=13$~TeV, compared to a NLO-NRQCD calculation (left) and to lower energy measurements from ALICE.}
\end{figure}

ALICE has also measured the $\psiprime$-to-$\jpsi$ inclusive cross section ratio as a function of $\pt$ and $y$. In figure~\ref{fig_ratio}, this ratio is compared to a NLO NRQCD calculation performed by the same group as in figure~\ref{fig_jpsi_yqma_y} (left) and to measurements performed at lower center of mass energy, namely $\ss=7$~\cite{Abelev:2014qha} and $8$~TeV~\cite{Adam:2015rta} (right). The motivation for measuring this quantity is that many sources of uncertainty cancel when forming the $\psiprime$-to-$\jpsi$ ratio, this both for the data and for the calculation. With these reduced uncertainties, it is observed that the NRQCD calculation slightly overestimates the ratio, whereas it was able to reasonably reproduce the $\psiprime$ and $\jpsi$ cross sections separately. In principle, the contribution from non-prompt $\psiprime$ and $\jpsi$ should be added to the calculation before comparing to the data. It was checked however that the impact of doing so is small, even at high $\pt$,  because the said contribution enters both the numerator and the denominator, and largely cancels.

Finally, considering the right panel of figure~\ref{fig_ratio}, no strong energy dependence is observed for the $\jpsi$-to-$\psiprime$ cross section ratio, for energies ranging from $\ss=7$~TeV to $13$~TeV. This also is consistent with models.

To summarize, ALICE has measured the inclusive $\jpsi$ and $\psiprime$ production cross sections at forward rapidity in pp collisions at $\ss=13$~TeV. This results complements the measurements performed by ALICE in the same rapidity range at $\ss=2.76$~\cite{Abelev:2012kr}, $5$~\cite{Adam:2016rdg}, $7$~\cite{Abelev:2014qha} and $8$ TeV~\cite{Abelev:2014qha}. The $\jpsi$ cross sections are in excellent agreement with the ones measured by LHCb at the same energy and in the same rapidity range, and extend the $\pt$ reach of the measurement from $14$ to $30$~GeV/$c$. For $\psiprime$, these cross sections are the first ones available at this collision energy and in this rapidity range. Cross section measurements are well described by NRQCD calculations provided that the non-prompt contribution is properly accounted for, for instance using FONLL. The same NRQCD calculation however slightly overestimates the $\psiprime$-to-$\jpsi$ ratio. When comparing this ratio to lower energy measurements, no strong energy dependence is observed.

\section*{References}
\bibliographystyle{iopart-num}
\bibliography{Hugo_Pereira_proceedings_SQM2016.bib}{}
%\end{thebibliography}

\end{document}